\documentclass[superscriptaddress,aps,prd,twocolumn,showpacs,nofootinbib]{revtex4-1}
\usepackage{etex}
\usepackage{amsmath,amssymb,amsthm}
\usepackage[utf8x]{inputenc}
\usepackage[colorlinks=true,citecolor=blue,urlcolor=blue]{hyperref}
\usepackage[pdftex]{graphicx}  
\usepackage{verbatim}   
\usepackage{xcolor}      
\usepackage{subfigure}  
\usepackage{hyperref}   
\usepackage{mathrsfs}
\usepackage{booktabs}
\usepackage[font=footnotesize,labelfont=bf]{caption} 
\usepackage{wrapfig}
\begin{document}
\title{Enhanced power of gravitational waves and rapid coalescence of black hole binaries through dark energy accretion}
\author{Arnab Sarkar}         \email{arnabsarkar@bose.res.in, arnab.sarkar14@gmail.com}
\affiliation{Department of Astrophysics and High Energy Physics, 
S. N. Bose National Centre for Basic Sciences, JD Block, Sector III, 
Salt lake city, Kolkata-700106, India}
\author{Amna Ali }  \email{amnaalig@gmail.com}
\affiliation{RsRL, Dubai, United Arab Emirates}
\author{K. Rajesh Nayak}   \email{rajesh@iiserkol.ac.in}
\affiliation{Indian Institute of Science Education and Research (IISER), Mohanpur, West Bengal-741246, India}
\author{A. S. Majumdar}   \email{archan@bose.res.in}
\affiliation{Department of Astrophysics and High Energy Physics, 
S. N. Bose National Centre for Basic Sciences, JD Block, Sector III, 
Salt lake city, Kolkata-700106, India}
\date{\today}
\begin{abstract}
\begin{center}
\textbf{Abstract}
\end{center}
\begin{small}
We consider the accretion of dark energy by constituent black holes in
binary formations during the present epoch of the Universe. In the context of
an observationally consistent dark energy model, we evaluate the growth
of black holes' masses due to accretion. We show that accretion leads to
faster circularization of the binary orbits. We compute the average power of 
the gravitational waves emitted from binaries, which exhibits a considerable
enhancement  due to the effect of growth of masses as a result of accretion. 
This in turn, leads to a 
significant reduction of the coalescence time of the binaries. We present
examples pertaining to various choices of the initial masses of the black
holes in the stellar mass range and above, in order to clearly establish
a possible observational signature of dark energy in the emerging era of gravitational wave astronomy.     
\end{small}
\end{abstract}
\maketitle
\begin{small}
\section{Introduction}
After the first direct detection of gravitational waves from a merging binary of black holes by aLIGO \cite{LIGO-VIRGO-1},  and subsequent series of detections from similar sources \cite{LIGO-VIRGO-2, LIGO-VIRGO-3, LIGO-VIRGO-4}, a new era in observational astronomy has begun. Besides binaries of compact objects in bounded orbits, there are various other mechanisms of production  of gravitational waves from a wide varieties of sources, such as nearby fly-pass of two compact objects in unbounded orbits \cite{Turner}, gravitational collapse of sufficiently massive stars \cite{Fryer_et_al}, cosmological phase transitions \cite{Hogan_et_al, Witten_et_al}, breaking of cosmic strings \cite{Battye_et_al, Leblond_et_al}, inflation and pre-heating \cite{Bellido_et_al, Dufaux_et_al} etc.. However, till date the observations by the aLIGO and VIRGO detectors have been carried out from mainly one type of sources, which are the binaries of compact objects, {\it viz}., black holes and neutron stars.
 
Gravitational wave observations have been used to estimate and constrain  various astrophysical and cosmological parameters associated with the generation and propagation of these gravitational waves. Among these, some important ones worth mentioning are: (i) estimating the Hubble parameter \cite{Abbott_et_al-1, Abbott_et_al-2}, (ii) constraining a large class of cosmological scalar-tensor theories \cite{Sakstein_et_al, Ezquiaga_et_al}, (iii) constraining the mass of gravitons for bimetric-gravity theories \cite{Baker_et_al}, (iv) investigating the state of matter inside a neutron star \cite{Pratten_et_al}, (v) constraining higher-dimensional theories \cite{Corman_et_al}, and there are several others. Attempts to constrain dark energy, responsible for the accelerated expansion of the late Universe \cite{Peebles}, have been made indirectly using gravitational wave observations, either through the estimation of the Hubble parameter, or through constraining cosmological scalar-tensor theories. 

Late time acceleration of expansion of the Universe is one of the most intriguing discoveries of recent times, which was directly confirmed from supernovae Ia observations in 1998 \cite{Riess_et_al, Perlmutter_et_al} and was also supported by various indirect probes. Many theoretical approaches have been employed to explain the current  cosmic acceleration. The component of the Universe providing the required negative pressure for driving this accelerated expansion is generically called `dark energy' \cite{Sahni}. As normal matter (radiation, baryonic matter or cold dark matter) is gravitationally attractive, the standard lore is to assume the presence of a relativistic fluid which is repulsive in nature, as the dark energy candidate. The simplest candidate of dark energy is the cosmological constant $\Lambda$, which is mostly consistent with cosmological observations. However, it is plagued with conceptual problems, for example, fine-tuning and coincidence problems \cite{brax}, which are theoretical in nature. With a hope to address these problems, cosmologists have proposed mechanisms where the role of dark energy is played by a completely different component of the Universe, which may have a variable equation-of-state parameter. 
 
Many varieties of dark energy models have been proposed, theoretically studied and observationally constrained till now. There exist a wide class of scalar field models coupled to gravity. Among these, minimally coupled ones, called quintessence, in which cosmic acceleration is driven by the potential energy \cite{Caldwell_et_al, Zlatev_et_al}, are known to alleviate some of the problems of the cosmological constant. Scalar fields, in which the cosmic acceleration is driven by the kinetic energy, called `k-essence' \cite{Armendariz-Picon_et_al-3, Armendariz-Picon_et_al-1, Armendariz-Picon_et_al-2}, have also been studied, motivated from unification and quantum gravity scenarios. Such models may further yield a consistent picture of the complete evolution, starting from the early era inflation, the subsequent dark matter domination, and finally the late time acceleration \cite{asm2, asm3}. Other alternatives include random barotropic fluids with pre-determined forms of the equation-of-state parameter, such as the Chaplygin gas models \cite{Bento_et_al}, string theory motivated models \cite{Sen, Padmanabhan, Ali1} and braneworld models \cite{Dvali_et_al, asm1}. There also exist approaches without requiring additional fields\cite{Schwarz, Rasanen, Wiltshire, Ali2}. 

A major difference between the scalar field and other fluid models of dark energy with the $\Lambda$CDM model (and other approaches not requiring additional fields) is that the former type of dark energy is subjected to accretion by  the black holes present in the Universe. In fact, those back holes with surroundings containing insufficiently available other forms of matter-energy for accretion, would still accrete the scalar field dark energy, which is uniformly distributed throughout the Universe. Accretion of various types of dark energy by black holes has been a subject of theoretical interest for a considerable time \cite{Babichev_et_al-1, Cheng-Yi, Pepe_et_al, Babichev_et_al-2, Gao_et_al}.
On the basis of various works done till date, it is widely accepted that the mass of a black hole would increase due to steady spherical accretion if the equation-of-state parameter of the dark energy $w$ is $ > -1$. On the other hand, accretion would result in mass loss of a black hole, for phantom type dark energy with $w < -1$.
 
If dark energy exists in the Universe in a form which can be accreted by black holes, the result would not be limited to just the change of masses of the black holes. It is expected that  other phenomena associated with the black holes would also be influenced. The evolution of binaries formed with the black holes, the gravitational wave emitted from those binaries and coalescence of those binaries are some of the physical processes which get directly affected if the masses of the concerned black holes  are changing continuously instead of being constant. The efficacy of the above effects, in particular, whether the modified variation of gravitational energy of the binary system could be detectable via the rate of change of the orbital radius, has been a subject of debate \cite{Houghton_et_al, Enander_et_al} in the case of spherically symmetric accretion of dark energy \cite{Babichev_et_al-2}. 

In the present work our motivation is to explore the problem of associated modification of black hole binary parameters due to accretion in the context of a popular k-essence model of dark energy. Specifically, we consider a string theory \cite{string} inspired low energy effective action framework containing a dilaton scalar field \cite{string2}. The resultant k-essence dark energy scenario \cite{Piazza_et_al} is compatible with cosmological  observations \cite{Ohashi_et_al}. Here we study spherical accretion of the k-essence dilatonic ghost condensate dark energy by black holes. This falls within a class of models known as `ghost condensates' \cite{Arkani-Hamed_et_al}.
Considering binary formations of black holes in the early inspiral stage, we study main aspects of evolution of the orbits, due to continuous change of masses of the component black holes of such binaries, resulting from spherical accretion of the chosen model of k-essence dark energy. More specifically, we study the pace of shrinking of such an orbit and the average power of the gravitational wave emitted from the orbit in the course of its evolution, and perform 
quantitative comparisons of the differences with the case when the masses of component black holes are constant. We further investigate the modification in the time required to reach the coalescence stage of such a binary, in comparison with the constant mass case. 
 
The paper is organised as follows. In section \ref{Section-2}, we study the growth of black hole mass due to accretion of the chosen model of k-essence dark energy in the late Universe. In section \ref{Section-3}, we investigate the effect of growing masses of black holes on the evolution of binaries. We compute the rate of decrease of orbital radius after circularization of orbits, and the average power of the emitted gravitational waves. We compare these results with the case of binaries with black holes of constant masses without accretion. In section \ref{Section-4}, we estimate the reduction in the time required for reaching coalescence-stage by such binaries. We present our concluding remarks in section \ref{Section-5}.   
 
\section{Dark energy accretion by black holes in a k-essence model} \label{Section-2}

K-essence scalar fields are the dynamical dark energy models where the acceleration is driven by kinetic term in the scalar field Lagrangian. Among many k-essence models, we choose a particular string-inspired ghost condensate model, called `k-essence dilatonic ghost condensate', which can successfully describe the cosmological evolution, while simultaneously satisfies the necessary conditions of quantum stability and sound speed \cite{Piazza_et_al, Amendola_et_al}. This model has also found to be observationally consistent \cite{Ohashi_et_al}. 
\\
The condition on sound speed for any scalar field dark energy model is simply that the sound speed can not exceed the speed of light in vacuum (c) i.e. it can not have super-luminal speed. 
In this regard, it is worthwhile to mention that the sound speed makes an important difference between quintessence and k-essence models. While for standard quintessence models, with canonical scalar fields, the sound speed is always equal to the speed of light ; for the k-essence models it is not so. This fact of having varying sound speeds through the  cosmic evolution gives various ways of distinguishing different k-essence models from one-another and from the quintessence models \cite{Erickson_et_al}.
 In fact this difference of sound speed of k-essence models with the quintessence models is one of the main reasons, for which we have chosen a k-essence model for this study. 
 \par
The action of k-essence scalar field $ \varphi$, along with non-relativistic matter and radiation, can be generally written as \cite{Armendariz-Picon_et_al-3} : 
\begin{equation} \label{2.0a}
S = \int d^{4}x \sqrt{-g} \left\lbrace \frac{1}{2 \kappa^{2}} R + \mathcal{L}(\varphi, X) \right\rbrace + S_m  \, , 
\end{equation}  
where $\kappa = (8 \pi G/3)^{1/2} $, $R$ is the Ricci-scalar and $\mathcal{L}$ is a function of the k-essence scalar field $\varphi $ and its kinetic energy $X = -(1/2)g^{\mu \nu} \partial_{\mu} \varphi  \partial_{\nu} \varphi $. $S_m $ is the action contributed from the non-relativistic matter and radiation. In case of the specific model considered here, the Lagrangian density is given by \cite{Piazza_et_al, Amendola_et_al}: 
\begin{equation} \label{2.0b}
\mathcal{L} = - X + e^{\kappa \lambda \varphi} \frac{X^{2}}{M^4} \, , 
\end{equation}
where $ M $ is a constant having the dimension of mass and $\lambda $ is a constant dimensionless parameter, which is set according to stability conditions.

\begin{figure}[h]
\includegraphics[width=8.3cm]{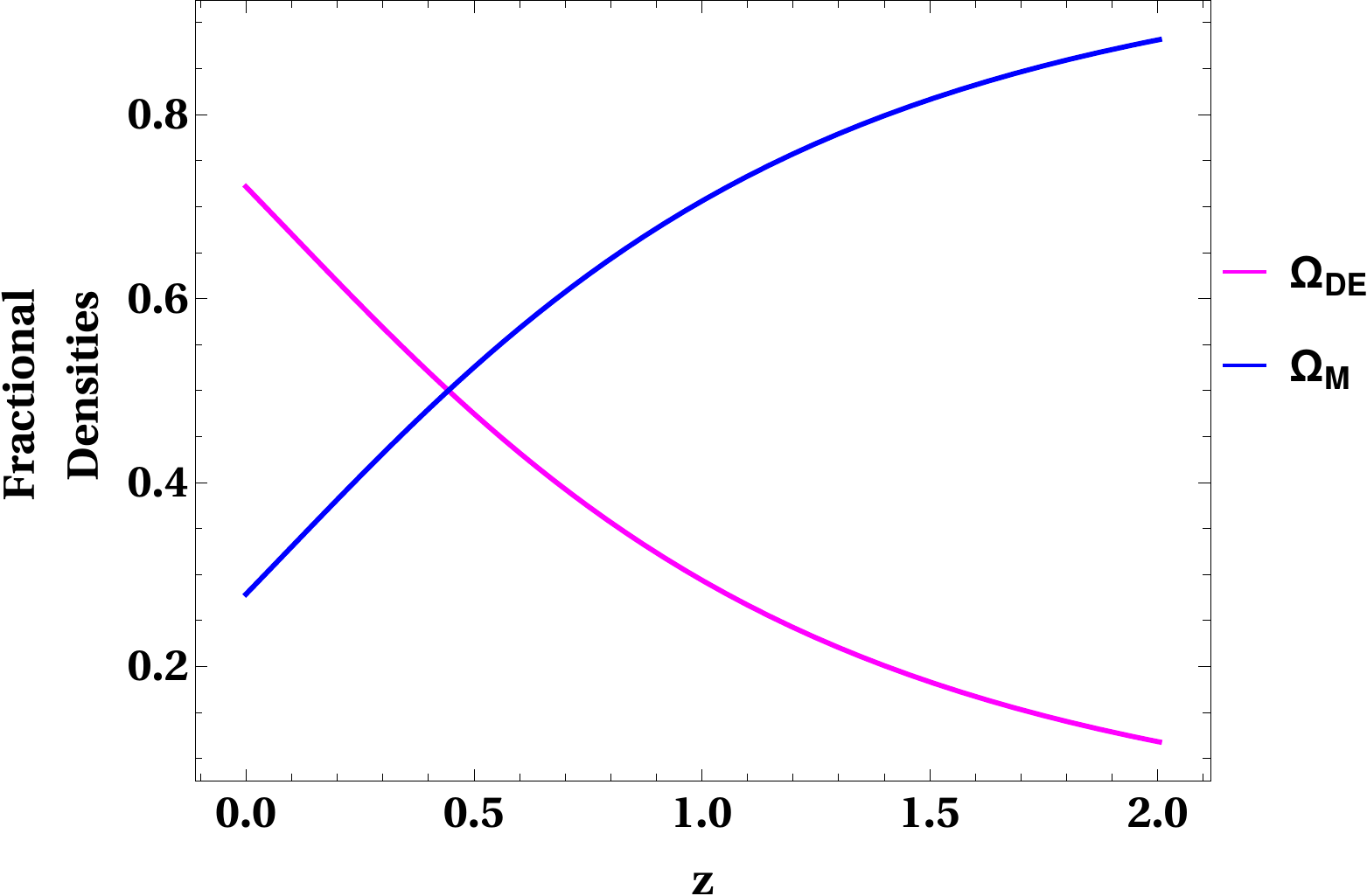}
\caption{Evolution of the fractional densities of k-essence dark energy denoted by $ \Omega_{DE} \equiv  \Omega_{\varphi}$ and non-relativistic matter denoted by $\Omega_{M}$,  w.r.t. redshift $z$. The fractional density of radiation $\Omega_{R}$  is negligible in this era of the Universe. }
\label{frac_den}
\end{figure}

 The set of equations governing the cosmological dynamics of this k-essence model can be conveniently written in terms of three dimensionless parameters: \cite{Piazza_et_al, Amendola_et_al} :
\begin{equation} \label{2.1}
x_1 = \frac{\kappa \dot{\varphi}}{\sqrt{6} H} \, , x_{2} = \frac{\varphi^{2} \, e^{\kappa \lambda \varphi}}{2 M^{4}} \, , x_{3} = \frac{\kappa \sqrt{\rho_{r}}}{\sqrt{3} H } \, , 
\end{equation}   
where $H$ is the Hubble-parameter and $ \rho_r$ is the density of radiation in the Universe. With these dimentionless parameters $x_1, x_2$ and $x_3$, the evolution equations can be cast in the following autonomous form:
\begin{equation}\label{2.2} 
\begin{aligned}
\frac{dx_1}{dN} = -\frac{x_1}{2}\frac{6(2 x_2 - 1 ) + 3 \sqrt{6} \lambda x_{1} x_{2} }{ (6 x_2 - 1)}
 \\
 + \frac{x_1}{2} (3- 3 x_1^2 + 3 x_1^2 x_2  + x_3 )   \, , 
\end{aligned}
\end{equation}
\begin{equation}
\frac{dx_2}{dN} = x_2 \frac{3 x_2 ( 4 - \sqrt{6} \lambda x_1 ) - \sqrt{6}(\sqrt{6} - \lambda x_1) }{1 - 6 x_2} \, ,   \label{2.3} 
\end{equation}
\begin{equation}
\frac{dx_3}{dN} = - \frac{3}{2} ( -x_1^2 + x_1^2 x_2 + \frac{x_3^2}{3} + 1 ) \, .  \label{2.4}  
\end{equation}
where $N = ln(a) = - ln (1+z)$ ; while $a$ and $z$ are respectively the scale-factor and the redshift. $N$ is generally called \textit{e-foldings}. 
The advantage of these autonomous equations and the dimensionless parameters is that, these are easier to solve numerically, and various important cosmological quantities can be given in terms of these dimensionless parameters  
$x_{1}, \, x_{2} $ and $ x_{3} $ {\it viz.} \cite{Piazza_et_al, Amendola_et_al},
\begin{eqnarray}
w_{eff} = -1 - \frac{2 \dot{H}}{3 H^{2}}  =  - x_1^2 + x_1^2 x_2 + \frac{1}{3} x_3^2 \, , \label{2.5} \\
w_{\varphi}  = \frac{1 - x_2}{1 - 3 x_2} \, , \label{2.6} \\
c_{s}^{2} =\frac{ 2 x_2 - 1 }{6 x_2 - 1} \, , \label{2.7} \\  
\Omega_{\varphi} = - x_1^2 + 3 x_1^2 x_2  \, ,\label{2.8} \\
\Omega_{R} = x_3^2                      \, ,  \label{2.9} \\
\Omega_{M} =  1 + x_1^2 - 3 x_1^2 x_2 - x_3^2 \, \label{2.10} , 
\end{eqnarray}
where $w_{eff}$ and $w_{\varphi}$ are respectively the effective equation-of-state parameter and the equation-of-state parameter of the k-essence  model. $ c_{s}$ is the sound speed of the k-essence model. $\Omega_{\varphi} , \Omega_{M} $ and $ \Omega_{R} $ are respectively the fractional densities of the dark energy, non-relativistic matter and radiation in the Universe.  
In the Fig. \ref{frac_den}, the evolutions of the fractional densities $ \Omega_{\varphi} $ and $ \Omega_{M} $ have  been depicted w.r.t. redshift $z$, for a certain period in the late Universe, with the initial conditions taken as $x_1 = 6.0 \times 10^{−11} \, ,
x_2  = 0.5 + (1.0 \times 10^{−9})$, and $ x_3 = 0.999$ at redshift $z \approx 10^{6.218} $  and the value of $\lambda = 0.2 $ \cite{Amendola_et_al}. 

\begin{figure}[h]
\includegraphics[width=6.5cm]{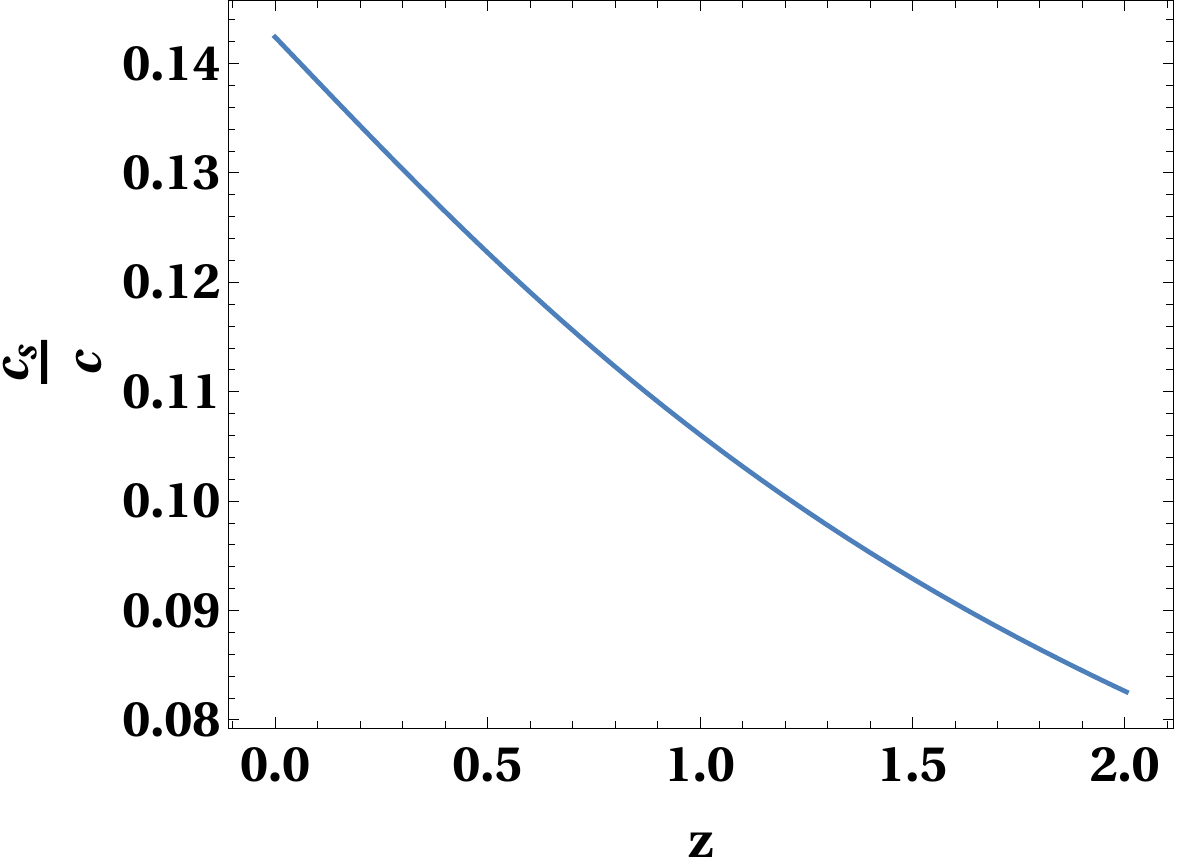}
\caption{The variation of the sound speed $c_s $ of the k-essence model w.r.t. redshift $z$. }
\label{Sound_speed}
\end{figure}

One can obtain simple equations for the Hubble-parameter $H$ and the time $ t $ in terms of the e-foldings $N$ , using  equation \ref{2.5} : 
\begin{equation} \label{2.11}
 \frac{1}{H^{2}}\frac{dH}{dt} = -\frac{3}{2} ( - x_1^2 + x_1^2 x_2 + \frac{1}{3} x_3^2 + 1 ) \, . 
\end{equation}
As $\frac{dN}{dt} = H $, the L.H.S. of the equation \ref{2.11} can be expressed as : 
\begin{equation*}
  \frac{1}{H^{2}}\frac{dH}{dt} =  \frac{1}{H^{2}}\frac{dH}{dN} \frac{dN}{dt} = \frac{1}{H} \frac{dH}{dN}  \, .  
\end{equation*}
Denoting $ h = ln \, H $, we get : 
\begin{equation} \label{2.12}
\frac{dh}{dN} = \frac{d}{dN} (ln \, H ) = \frac{1}{H} \frac{dH}{dN} \, . 
\end{equation}
Using the equation \ref{2.12}, we can write the equation \ref{2.11} as: 
\begin{equation} \label{2.13}
\frac{dh}{dN} = -\frac{3}{2} ( - x_1^2 + x_1^2 x_2 + \frac{1}{3} x_3^2 + 1 ) \, , 
\end{equation}
by solving which we can get $h$ and consequently $H = e^{h}$. 
After solving equation \ref{2.13} for the Hubble-parameter, we can simply solve the equation:
\begin{equation} \label{2.14}
\frac{dt}{dN} = \frac{1}{H} \, , 
\end{equation} 
to get the time $t$. 
Also, for the scale factor $a$, we have the equation: 
\begin{equation} \label{2.15}
\frac{da}{dN} = a \, . 
\end{equation}
Solving the above equation \ref{2.15} one can obtain the scale factor $a$ and corresponding redshift from the relation $1+z = 1/a$.  
We numerically solve the set of equations \ref{2.2}, \ref{2.3}, \ref{2.4} along with the equations \ref{2.13}, \ref{2.14} and  \ref{2.15}, with appropriate initial conditions.

We now consider accretion of dark energy by black holes in the context
of the above dark energy model. It may be noted here that the rate of
accretion is affected by the sound speed of the ambient fluid. The surface
of accretion is defined by the black hole horizon if there is no critical point outside the horizon \cite{Babichev_et_al-2}. The fluid being accreted by a black hole has the critical point, if its speed increases from subsonic to transonic values. From the historical development of spherical accretion by black holes starting from the pioneering work by Bondi \cite{Bondi}, it is evident that if a black hole moves through the ambient medium with a speed much lesser than speed of light in vacuum ($c$), and the medium, considered as a perfect fluid, has a sound speed less than $c$, then the accretion radius would be larger than the black hole horizon i.e. the Schwarzschild radius  \cite{Petrich_et_al}.

In the above scenario where the sound speed of the k-essence model lies in the range $0 < c_s/ c < 1 $, the time-rate of change of mass of a black hole spherically accreting the k-essence dark energy is obtained by using the
accretion radius $r_{a} = m / (v_{rel}^{2} + c_s^2)$ which defines the relevant 
surface of accretion \cite{Petrich_et_al}. Hence, the rate of accretion is given by
\begin{equation}  \label{2.16}
\frac{dm}{dt} = 4 \pi \mathscr{A}  \frac{ G^{2}m^{2}}{(v_{rel}^{2} + c_{s}^{2})^{3/2}} (1+w_{\varphi}) \rho_{\varphi} \, , 
\end{equation}
where  ${\displaystyle c_{s} =  \Big(\frac{dP}{d\rho}   \Big)^{1/2}}  $ and $ v_{rel} $ is the relative speed of the black hole  with respect to the ambient cosmic fluid being accreted. $ \rho_{\varphi} $ is the background density of the k-essence dark energy, $w_{\varphi}$ is the equation-of-state parameter of the k-essence dark-energy. $\mathscr{A}$ is a proportionality-factor that can be taken to be of the order of $\sim 1 $ \cite{Petrich_et_al}. Note that for $v_{rel} << c_{s} $ and $c_{s} \sim c $,  the equation \ref{2.16} leads to the rate of change of mass derived in Ref. \cite{Babichev_et_al-2}.

For the present analysis we consider that the relative speed of the moving black hole with respect to the ambient cosmic fluid, {\it viz.}, the k-essence  dark energy, is negligible in comparison to the sound speed of the  k-essence  
model, {\it i.e.}, $ v_{rel} << c_{s} $. This  is valid for most of the black holes in late  Universe, as it can be seen that in the dark energy dominated Universe, the sound speed of the chosen k-essence model is of the order of $\sim 0.1c $ (see Fig.\ref{Sound_speed}). So,  in the denominator on the R.H.S. of equation \ref{2.16}, $ v_{rel}^{2} $ can be neglected in comparison to $ c_{s}^{2} $. Thereby, the time-rate of change of mass of a black hole due to spherical accretion of the k-essence dark energy is given by 
\begin{equation} \label{2.16ii}
\frac{dm}{dt} = 4 \pi \mathscr{A}  \frac{ G^{2}m^{2}}{ c_{s}^{3}} (1+w_{\varphi}) \rho_{\varphi} \, ,
\end{equation}
where $2GM/c_s^2$ is the effective accretion radius, sometimes referred as the `Bondi radius'.

\begin{figure}[h]
\includegraphics[width=7.4cm]{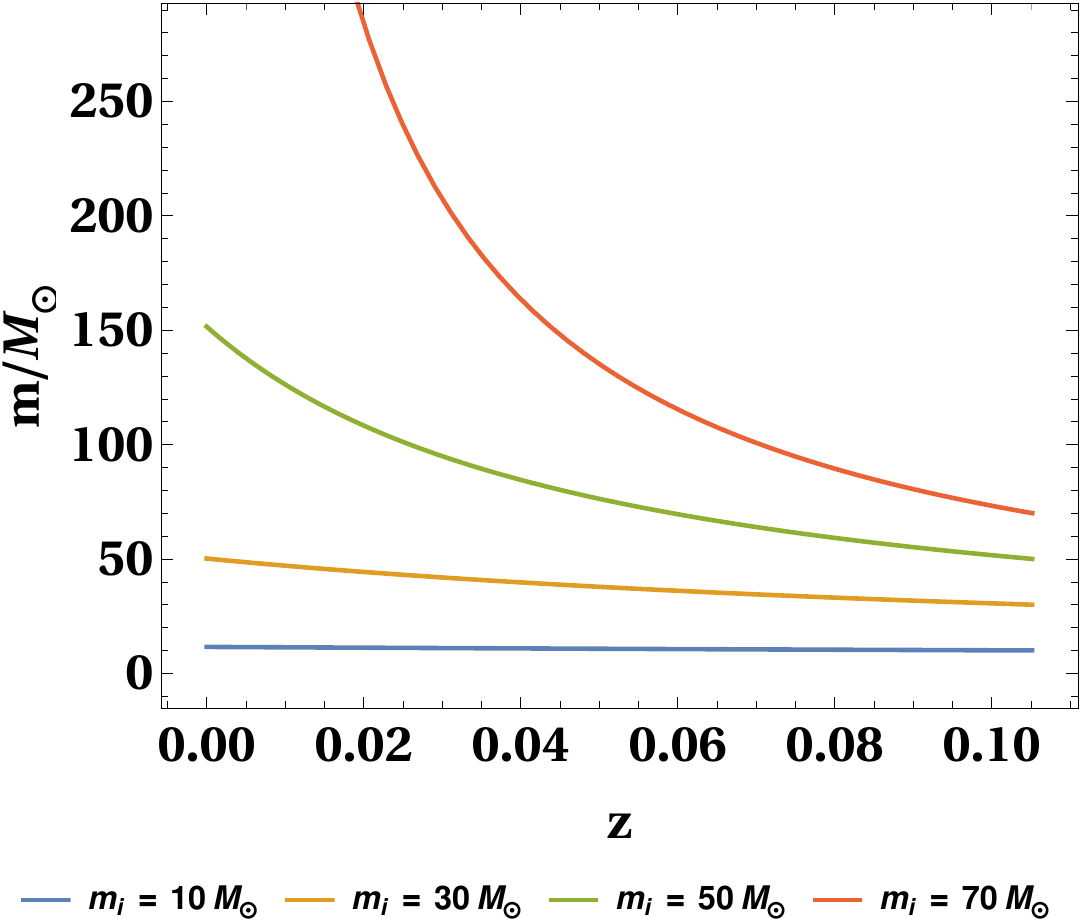}
\caption{ Growth of mass of black holes with various initial masses due
to accretion of the k-essence dark energy w.r.t. redshift $z$.  }
\label{Evol_massBHs}
\end{figure}

We can determine the time-rate of change of mass of a  black hole due to spherical accretion of the chosen k-essence dark energy using equation \ref{2.16ii},
where $ \rho_{\varphi} = \rho_{T} \Omega_{\varphi} $, and $\rho_{T} =( 3/ 8 \pi G ) H^{2} $ is obtained by solving the equation \ref{2.13} for getting the Hubble-parameter $H$.  
We depict the evolution of masses of black holes, due to accretion of the chosen k-essence dark energy, with four different initial masses taken as $ 10, 30, 50 $ and $70$ times of the Solar-mass ($M_\odot$) respectively, with respect to the redshift $z$ in Fig.\ref{Evol_massBHs}. 
It can be seen from the Fig.\ref{Evol_massBHs} that the amount of growth in masses of the black holes due to the dark energy accretion increases with the increase in their initial masses. It may be noted that ordinary stellar-mass black holes which are usually observed by  electromagnetic signals emitted by various type of astrophysical mechanisms  generally have masses in the range $ 5-20 \, M_\odot $. However, aLIGO and VIGO have detected gravitational waves from mergers of binaries  with  component black holes having masses from 30 $M_\odot$ to as large as 80 $M_\odot$ \cite{catalog-1, catalog-2}. 
It is quite evident from the Fig.\ref{Evol_massBHs} that stellar-mass black holes, having mass in the range $ 5-20 \, M_\odot $, can grow to  heavier ones by means of continuous spherical accretion of similar type of dark energy.

\section{ Power of  gravitational waves emitted from binaries  } \label{Section-3}

The instantaneous power of gravitational radiation due to the orbital motion of two black holes of masses $m_1 $ and $m_2$  in the quadrupole-approximation is given by \cite{Peters_et_al, M.Turner} :
\begin{equation} \label{1.1}
\begin{aligned}
\mathcal{P}(t)=
 \frac{8}{15} \frac{G^{4}}{c^{5}} \frac{M (m_{1} m_{2})^{2} }{(r_{min}(1+e))^{5}} (1+ e \, Cos \phi)^{4} 
 \\
 \left\lbrace  e^{2} Sin^{2}\phi + 12 (1+ e \, Cos \phi)^{2} \right\rbrace \, ,
\end{aligned}
\end{equation}
where $M= m_{1} + m_{2} $, `$e$' is the eccentricity of the orbit, $\phi$ is the angular position of the reduced mass $\mu $ on the plane of the orbit in a polar-coordinate system $(r, \phi)$ with origin at the center-of-mass, and $r_{min}$ is the radial distance of closest approach. 
In the present  case, the masses are continuously changing due to accretion of dark energy. Due to this continuous time-variation of the masses of the black holes, two extra terms (having single and double time-derivatives of the masses) arise along with the main term in the amplitude of gravitational radiation \cite{arnab1}. However, these terms are negligible in comparison to the main term in this case. Hence, the equation \ref{1.1} needs to be considered here with time-dependent masses. 

When the orbit is bounded, the total energy carried away by the gravitational radiation due to the relative motion of the system of two black holes within one complete cycle or time-period, is given by 
\begin{equation} \label{1.2}
\Delta\mathcal{E} = \int^{T}_{0} \mathcal{P}(t) dt = \int^{2\pi}_{0} \mathcal{P}(t) \frac{dt}{d \phi} d\phi \, . 
\end{equation}    
It is known that energy of gravitational waves is well-defined when the average of the energy over several time-periods of the wave is taken. Also, a compact object in a Keplarian elliptical orbit emits gravitational waves with frequencies, which are integral multiples of the frequency $ \omega_0 = \left( G M/ a^{3} \right)^{1/2} $, where $a$ is the semi-major axis of the elliptical orbit. Hence, the period of the gravitational waves emitted due to this orbital-motion, is a fraction of the orbital-period. Therefore, a well-defined version of the power of the emitted  gravitational waves is the average of the power taken over one period of the orbit.  
The average of the power $P_{avg}$ over one period of the orbit can be written as \cite{Peters_et_al} 
\begin{equation} \label{1.3}
P_{avg}(t) = \frac{1}{T} \int_{0}^{T}  \mathcal{P}(t) dt = \frac{32 G^{4}(m_1 m_2)^{2} M }{5 c^{5} a^{5}} f(e) \, , 
\end{equation}       
where the function $f(e)$ of eccentricity $e$ is given by : 
\begin{equation} \label{1.4}
f(e) = \frac{1}{(1-e^2)^{7/2}} \left( 1 + \frac{73}{24} e^2 + \frac{37}{96} e^4 \right) \, . 
\end{equation}
For a circular orbit $e=0$, thereby $f(e)$ becomes 1 and $a$ becomes the radius of the circular orbit.  
Note that in case of constant masses the eccentricity of the orbit changes only due to the emission of gravitational waves. However, in the present case since the masses of the black holes are continuously changing  through accretion of dark energy, the change of eccentricity would be due to two different effects: (i) growth of the masses via accretion, and (ii)  loss of energy and angular momentum carried away by gravitational waves. \footnote{The angular momentum of the system of two black holes is not affected due to spherical accretion of  dark energy because the  scalar-field dark energy model considered here does not contain angular momentum, and hence, cannot impart any angular momentum to the system.} 

\begin{figure}
\includegraphics[width=7.4cm]{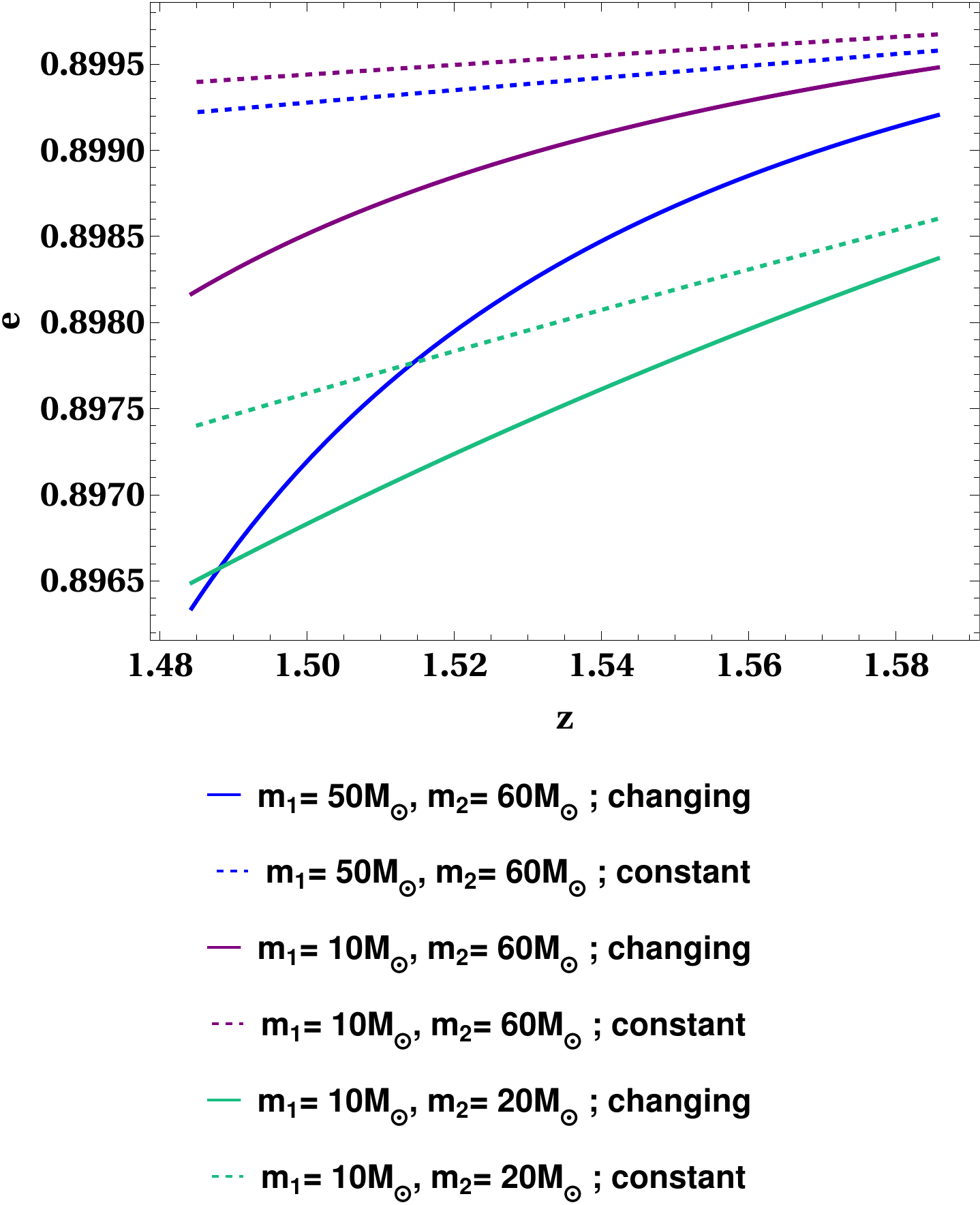}
\caption{Evolution of eccentricities of elliptical orbits, from the initial value 0.9, w.r.t. redshift $z$, for three different combinations of the initial masses of black holes, for two different cases, (i) growing masses and (ii) constant masses.  } \label{Eccen_changingmass}
\end{figure} 

Using the rate of change of energy and angular momentum of a binary of black holes in bounded orbit, the rate of change of the semi-major axis $a$ and eccentricity $e$ of the orbit can be obtained as \cite{Maggiore}, 
\begin{equation}
\frac{da}{dt} = - \frac{64}{5} \frac{G^{3} (m_1 m_2) M}{c^{5} a^{3}} \frac{1}{(1-e^2)^{7/2}} \left( 1 + \frac{73}{24} e^2 + \frac{37}{96} e^4 \right) \, ,   \label{1.5}
\end{equation}
\begin{center}
and
\end{center}
\begin{equation}
\frac{de}{dt} = - \frac{304}{15} \frac{G^{3}(m_1 m_2) M}{c^{5} a^{4}}  \frac{e}{(1-e^2)^{5/2}} \left( 1 + \frac{121}{304} e^2  \right) \,  .  \label{1.6} 
\end{equation}
It may be noted that the semi-major axis $a$ and eccentricity $e$, governed by the above equations \ref{1.5} and \ref{1.6}, are averages of these quantities over one period of the orbit, not their instantaneous values, as the corresponding equations of energy and angular momentum of the system, from which these are derived, govern their  averages over one period. This is quite evident from the fact that these equations \ref{1.5} and \ref{1.6} do not contain the phase-angle $\phi$.   
From the equation \ref{1.6} it follows that, if the eccentricity $e$ becomes zero($0$), then $\frac{de}{dt} = 0 $ implying $e =$ constant, {\it i.e.}, $e$ remains zero. Thereby, once the orbit becomes circular, it remains circular.   
\par
We solve these equations \ref{1.5} and \ref{1.6} numerically for different initial masses of the black holes forming  binaries and orbiting in elliptical orbits with initial eccentricity $e_i  = 0.9$. We choose the combinations of initial masses of the black holes forming the binaries to be $ 50,60M_\odot \, ; 10,60M_\odot $ and $10,20M_\odot $, respectively. 
\begin{widetext}
\begin{center}
\begin{figure}[h]
\includegraphics[width=12.8cm]{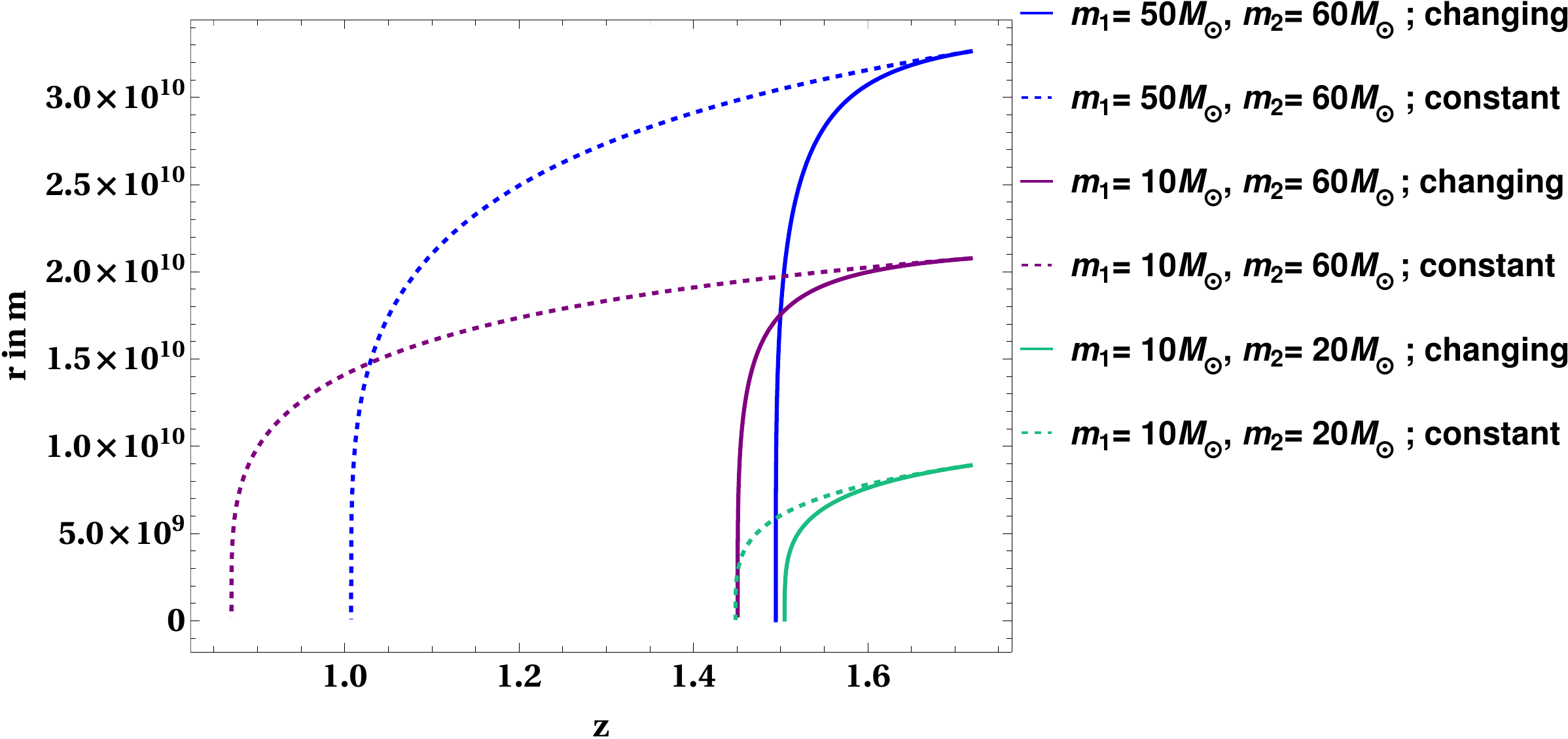}
\caption{Variation of the radius r of circular orbit of two black holes in binary formation w.r.t. redshift $z$, in three different combinations of initial masses and two cases {\it viz.}, (i) growing masses, and (ii) constant masses.} 
\label{Radius-vs-time}
\end{figure}
\end{center}
\end{widetext}  
 The initial semi-major axis $a_i $ of the elliptical orbit has been taken as $10^6$ times of the sum of their initial  Schwarzschild-radii,  {\it i.e.}, $a_i = 10^{6}  (2 GM_i/c^2 )$, ($M_i $ being the initial total-mass of the black holes) so that the Keplarian-approximation holds well. The time-period of the orbit is given by : $T = 2\pi/ \omega_0 $, where the angular-frequency $\omega_0$ is given by : 
$\omega_0 = \left( \frac{G M}{a^{3}} \right)^{1/2} \, $.  
\\ 
The fall of the eccentricities of the elliptical orbits of the binaries, for three different combinations of initial masses of the constituent black holes, is depicted in the Fig.\ref{Eccen_changingmass} (Only certain portions of the full evolution-profiles have been shown here so that the differences can be visualized clearly). It can be seen from Fig.\ref{Eccen_changingmass} that the eccentricities for the orbits of binaries, where the masses of the black holes are growing due to accretion of the chosen model of k-essence dark energy, drop faster than those where the masses are constant. Moreover, the eccentricities of binaries with larger mass black holes drop faster.\\
After the eccentricity vanishes, {\it i.e.}, circularization of the orbit is achieved, the rate of change of radius $r $ of the circular orbit is given by 
\begin{equation} \label{1.7}
\frac{dr}{dt} = - \frac{64}{5} \frac{G^{3} (m_1 m_2) M}{c^{5} r^{3}} \, . 
\end{equation}
Correspondingly, the average power of the emitted gravitational wave for the circular orbit becomes,
\begin{equation} \label{1.8}
P_{avg}(t) = \frac{32 G^{4}(m_1 m_2)^{2} M }{5 c^{5} r^{5}}  \, .  
\end{equation}

\begin{figure}[h]
\includegraphics[width=6.23cm]{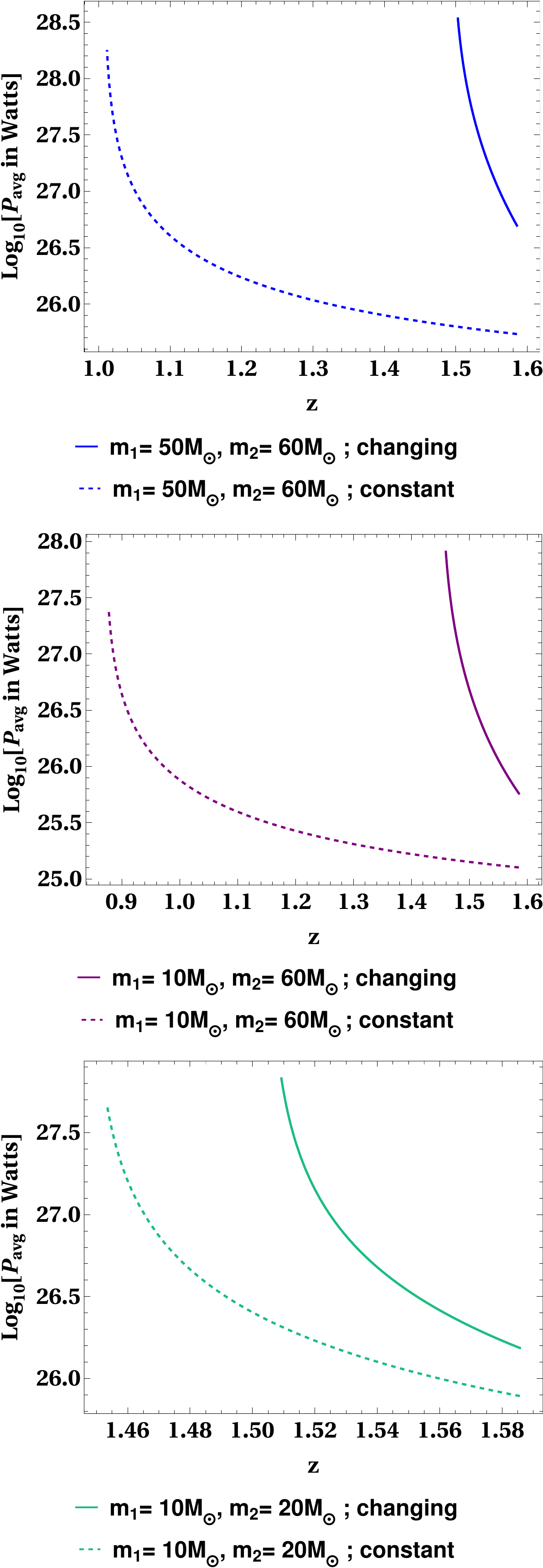}
\caption{Evolution of average power $P_{avg}$ of the gravitational wave emitted due to the orbital-motion of two black holes  in binary formation, w.r.t. redshift z, for three different combinations of initial masses and for two different (constant and changing mass) cases. A  range of the full evolution profiles have been shown for visual clarity.
 } \label{Power_comb}
\end{figure}

We first determine the patterns of shrinking of radius $r$, by solving the equation \ref{1.7}, for the circular orbits  in which two black holes of masses $m_1$ and $m_2$ are in binary formations, for two different cases {\it viz.}, (i) when the masses are changing due to spherical accretion of dark energy described by our chosen model and (ii) when the masses are constant, for three specified combinations of initial values of $m_1$ and $m_2$ for each of the cases. For this, we fix the initial radius for each of the cases to be $10^5$ times of the sum of the initial Schwarzschild radii of the black holes, {\it i.e.}, $r_i = 10^5( 2GM_i/c^2)$ (where $M_i$ stands for the initial total mass of the black holes). This choice for the initial radii of the circular orbits for each case is considered to study the comparative evolution with similar initial conditions. 
 The radii of the circular orbits for three different combinations of initial values of masses $m_1$ and $m_2$, and for two different cases, as mentioned above, are plotted w.r.t. redshift $z$ in Fig.\ref{Radius-vs-time}. It can be seen from the Fig.\ref{Radius-vs-time} that, with the increasing difference in the masses and increasing total masses of the component black holes of the binaries, the difference in rate of shrinking of the radii of the circular orbits increases. 
 
We next study the variation of the average power $P_{avg}(t)$ with the evolution of the circular orbits for each of the cases. We depict the variation of the average power  w.r.t. the red-shift for each of the cases in Fig.\ref{Power_comb}.  
It can be seen from the plots in Fig.\ref{Power_comb} that within the same interval, the average power of the emitted gravitational wave grows significantly higher for the binaries of black holes with growing masses, in comparison with the case when the masses of the black holes are constant. 
The average power of the emitted gravitational wave in case of evolving masses of black holes in the binaries, grows faster in comparison to the case of constant masses of the black holes. A certain amount of increase of the masses of the black holes of binaries results in more amplification of the average power, because of the fact that the average power of the emitted gravitational waves is proportional to the quantity $\mu^{2}M^{3}$ ($\mu$ being the reduced-mass of the black holes forming the binary). So, a small increment in the masses of the black holes results in a comparatively greater increase in the average power of the emitted gravitational waves. Moreover, the faster shrinking of the radius of circular orbit for increasing masses of the black holes, in comparison to the case of constant masses,  also contributes to the faster growth of the average power $P_{avg}(t)$ in the former case, as it is proportionl to $r^{-5}$.


\section{Reduced coalescence time } \label{Section-4}

From the previous analysis we have seen that as the masses of the black holes forming the binary increases due to accretion of dark energy, the average power of the emitted gravitational waves becomes significantly higher with the evolution of the orbit, in comparison to the case of constancy of masses when there is no accretion. Since the power of the emitted gravitational wave increases with time, the binary loses energy faster and shrinks more rapidly. As a result, the time taken by a binary to coalesce is shorter when the black holes' masses are growing, than for the case of constant masses of the black holes. Let us now estimate the decrease in coalescence time-interval of a binary, due to increasing masses of the component black holes that are spherically accreting the chosen model of k-essence dark energy. 

For a binary constituted with black holes of constant masses, the rate of loss of energy by the binary is equal to the power of the emitted gravitational waves, {\it i.e.}, $ P_{avg} = -dE_{avg}/dt $, where $E_{avg}$ is the average energy of the binary. Using the expression of average power from equation \ref{1.8},  the time-evolution of the frequency of gravitational waves ($f_{gw}$) emitted from the binary is given by  \cite{Maggiore}, 
\begin{equation}  \label{3.1}
\frac{d f_{gw}}{dt} = \frac{96}{5} \pi^{8/3} \left( \frac{G \mathcal{M}}{c^{3}} \right)^{5/3} f_{gw}^{11/3} \, , 
\end{equation}    
where $\mathcal{M} = (m_1 m_2)^{3/5}/M^{1/5} $ is the chirp-mass of the binary. For the case of constant masses of the component black holes of the binary, the solution of the above  equation \ref{3.1} can be written as \cite{Maggiore}: 
\begin{equation} \label{3.2}
f_{gw} = \frac{1}{\pi} \left( \frac{5}{256} \frac{1}{\tau} \right)^{3/8} \left( \frac{G \mathcal{M}}{c^3} \right)^{-5/8} \, ,
\end{equation}   
where $t_c$ is the time of coalescence of the binary and $\tau = t_c - t $ is the time-interval required by the binary to reach coalescence, from any stage of its evolution at an arbitrary time $t$. 
Using the equation \ref{3.2} evaluated at an initial time $t_i$ and the relation $\omega_{s \, i}^{2} = (G M/r_i^3) $, where $\omega_{s \, i}$ is the initial source-frequency, it can be shown that the time-interval $\tau_i = t_c - t_i$ required by the binary to reach the coalescence stage from initial instant, is related to the initial radius of the circular-orbit $r_i$, as \cite{Maggiore} 
\begin{equation} \label{3.4}
\tau_i = \frac{5}{256} \frac{c^5 r_i^4 }{G^3 M m_1 m_2 } \, . 
\end{equation}

Now, for the case when the masses of the black holes are changing, the
counterpart of the equation \ref{3.2} valid in the present case is given by
\begin{equation} \label{3.5}
f_{gw}^{-8/3} =  \frac{256}{5} \pi^{8/3} \left( \frac{G}{c^{3}} \right)^{5/3} \int_{t}^{t_c} \mathcal{M}^{5/3} dt  \, .
\end{equation} 
Making a change of variable from $t$ to $\tau$ in the integral $\int_{t}^{t_c} \mathcal{M}^{5/3} dt$, we can write: 
\begin{equation} \label{3.6}
\int_{t}^{t_c} \mathcal{M}^{5/3} dt =  \int_{0}^{\tau} \mathcal{M}^{5/3} d\tau \, . 
\end{equation}  
In following  the suffix `$i$'  denotes the corresponding initial value of the quantity at initial time $t'_i$.
 Next, as was done for the case of constant masses, here also we evaluate the equation \ref{3.5} at initial time $t'_i$ and use the relation $\omega_{s \, i}^{2} = (G M_i/r_i^3) $.
It follows that $t'_i$ or alternatively $\tau'_i$ satisfies the equation :
\begin{equation} \label{3.7}
 \frac{ {\displaystyle\int_{0}^{\tau'_i}  \mathcal{M}^{5/3} d\tau  }}{\mathcal{M}_i^{5/3}} = \frac{5}{256} \frac{c^5 r_i^4 }{G^3 M_i m_{1i} m_{2i} } \, ,
\end{equation}
as the counterpart of the equation \ref{3.4} for the case of changing masses of black holes in binaries, due to dark energy accretion.
The equations \ref{3.4} and \ref{3.7} provide the values of the coalescence time-intervals for the two different cases, {\it viz.}, constant mass and varying mass of the black holes, respectively. Therefore, when the initial masses of the component black holes and initial radii of orbits are same for both the cases, the difference of the coalescence-time intervals corresponding to the two cases is given by, 
\begin{equation} \label{3.8}
 \Delta \tau_i =  \tau'_i - \tau_i . 
\end{equation} 
 
In order to perform a comparative estimate of the reduction in coalescence time-intervals due to dark energy accretion  for the examples of binaries studied in the present work, we fix the initial time to be same for both the cases (pertaining to constant and changing masses), and evaluate the corresponding times of coalescence.
We obtain the times of coalescence for binaries of the three different combinations of initial black hole masses considered by us in the previous section, for studying the evolution of eccentricities of the elliptical orbits, shrinking of radii of the circular orbits and the average power of emitted  gravitational waves. We choose the initial time at the e-folding value $N = -1$, or the corresponding redshift $z \approx 1.72$. 
\begin{widetext}
\begin{center}
\begin{tabular}{|c|c|c|c|}
\hline
\toprule
{Initial masses} & {10 and 20 $M_\odot$} & {10 and 60 $M_\odot$} & {50 and 60 $M_\odot$}  \\
\hline \hline
{Initial radius of orbit $r_i$ } & {$ 8.899 \times 10^9 \, m $}  &  {$ 20.764 \times 10^9 \, m $}  & { $ 32.628 \times 10^9 \, m $ } \\
\hline
{$t_c$ for constant masses} & {$ 4.817 \,\, Gy $} & {$ 6.956 \,\, Gy $} & {$ 6.329 \,\, Gy $} \\
\hline
{$t'_c$ for varying masses}  &  {$ 4.665 \,\, Gy $}  &  {$ 4.81 \,\, Gy $}  &  {$ 4.693 \,\, Gy $} \\
\hline
{$\tau_i $ for constant masses} & {$ 66.135 \times 10^7 \, y$} & {$ 2.8  \, Gy$} & {$ 2.173 \, Gy$} \\
\hline
{$\tau'_i $ for varying masses} & {$ 50.99 \times 10^7 \, y $} & {$ 0.655  \, Gy $} & {$ 0.537 \, Gy$} \\
\hline
{Decrease in Coalescence-time $\Delta \tau_i =  \tau'_i - \tau_i$} & {$ 15.14 \times 10^7 \, y $} & {$ 2.146  \, Gy $} & {$ 1.637 \, Gy $}  \\
\bottomrule
\hline
\end{tabular}
\captionof{table}{Reduction in coalescence time-intervals due to accretion of the chosen model of dark energy.} \label{Table-1}
\end{center}
\end{widetext}
For each of the cases, we set the initial radii $r_i$ of the circular orbits to be $10^5$ times of the sum of the initial Schwarzschild radii of the black holes. We display the decrease in coalescence time-intervals for these three different examples in the Table \ref{Table-1}.
\par
From the above analysis it is evident that for binaries consisting of black holes, having masses in the stellar-mass range and few-times greater than the stellar-mass black holes, specifically those from which several merging events have been detected by the aLIGO and VIRGO detectors, the time required for coalescence gets significantly reduced due to the increase in masses of the black holes caused by accretion of the chosen model of dark energy. The magnitude of reduction in the coalescence time-interval is $\sim 10^8 \,$ years, when both the component black holes are stellar-mass black holes (note the column for the combination of 10 and 20 $M_\odot$ in Table \ref{Table-1}). However, for larger mass black holes (having masses few times larger than stellar-mass ones) the effect of accretion of the dark energy is greater. For example, the coalescence time-interval gets reduced by $\sim 10^{9}$ years (see the third column in Table \ref{Table-1}), and even more if there is a significant difference in the initial masses of the constituent black holes (see the second column in Table \ref{Table-1}). Note though, that the magnitude of decrease in coalescence time-interval also depends on the initial radius of the circular orbit.


\section{Conclusions } \label{Section-5}

A variety of cosmological observations have revealed that the present Universe is undergoing a 
phase of accelerated expansion, and such observations lend support to dynamical dark energy models responsible for the present acceleration. 
The string theory inspired dilatonic scalar field model, chosen in this work, in which acceleration is driven by the scalar field kinetic energy \cite{Piazza_et_al}, seems to be observationally consistent \cite{Ohashi_et_al}. If dark energy exists in an accretable form, it is inevitable that the black holes existing in the present Universe would evolve by accreting it. This in turn, would have a natural imprint on the evolution of binaries constituted by the black holes, as we have shown in this work.

Specifically, we have studied the effect of growth of masses of black holes due to the spherical accretion of the chosen model of k-essence dark energy on several important parameters of binaries constituted by those black holes. We have investigated the effect of changing masses of the black holes on the evolution of the binaries and the average power of the emitted gravitational waves. 
We have found that accretion of the chosen model of k-essence dark energy leads to rapid circularization of binary orbits in comparison to the case of constancy of masses i.e. without accretion. Further, in comparison with the constant-mass case, the average power of gravitational waves increases significantly faster due to the increase in masses of the black holes.
Since the average power grows as $P_{avg} \propto \mu^2 M^3$, a comparatively small amount of increase in masses due to accretion leads to a much larger increment of the average power of emitted gravitational waves within the concerned time-scale. Finally, we have analysed how the effect of the increase in masses of the black holes leads to the reduction in the coalescence-time intervals of  black hole binaries in the stellar mass range and above.

 Our work establishes the fact that if dark energy is similar to scalar field models like the k-essence model  considered here, then it would result in reduced coalescence-time intervals of the binaries of black holes present in the current era of the Universe. The reduction in coalescence-time intervals means increased rate of coalescences. A possible upshot of the effect of accretion of dark energy by black holes in binary formations is that if this effect is observationally detectable in the new era of gravitational wave astronomy, it can lead to independent constraints on the equation-of-state parameter $w$ of the dark energy model. Such observations on local candidates are associated with much less noise compared to certain other dark energy observations involving all-sky surveys such as cosmic microwave background and baryon acoustic oscillations. Observations with aLIGO and VIRGO detectors should be useful in this regard, as we have demonstrated the significance of the effect for the binaries of black holes within mass-ranges, from which several merging events have been detected by these detectors.
  Moreover, upcoming observations using the planned futuristic detector LISA, may also be able to investigate the imprint of dark energy accretion on coalescence time-intervals for binaries of supermassive black holes formed during galaxy mergers or even extreme mass-ratio inspirals (EMRIs) also. 
\vspace{1cm}
\\
\textbf{Acknowledgements:} Arnab Sarkar thanks S. N. Bose National Centre for Basic Sciences, Kol-106, under Dept. of Science and Technology, Govt. of India, for funding through institute-fellowship.
\newpage

\end{small}

\begin{thebibliography}{99}
\bibitem{LIGO-VIRGO-1}
B. P. Abbott {\it et al.} (LIGO-VIRGO Collaboration), Phys. Rev. Lett. {\bf 116}, 061102 (2016),
\bibitem{LIGO-VIRGO-2}
B. P. Abbott {\it et al.} (LIGO-VIRGO Collaboration), Phys. Rev. Lett. {\bf 116}, 241103 (2016),
\bibitem{LIGO-VIRGO-3}
B. P. Abbott {\it et al.} (LIGO-VIRGO Collaboration), Phys. Rev. Lett. {\bf 118}, 221101 (2017).
\bibitem{LIGO-VIRGO-4}
B. P. Abbott {\it et al.} (LIGO-VIRGO Collaboration), Astrophys. J. Lett. {\bf 851}, L35 (2017).
\bibitem{Turner}
M. S. Turner, Astrophys. J. {\bf 216}, 610 (1977).
\bibitem{Fryer_et_al}
C. L. Fryer, K. C. B. New, Living Rev. Rel. {\bf 14}, 1 (2011).
\bibitem{Hogan_et_al} 
C. J. Hogan, Phys. Lett. B {\bf 133}, 172 (1983).
\bibitem{Witten_et_al}
E. Witten, Phys. Rev. D {\bf 30}, 272 (1984).
\bibitem{Battye_et_al}
R. A. Battye, R. R. Caldwell, E. P. S. Shellard, arXiv: astro-ph/9706013.
\bibitem{Leblond_et_al}
L. Leblond, B. Shlaer and X. Siemens, Phys. Rev. D {\bf 79}, 123519 (2009).
\bibitem{Bellido_et_al}
J. Garc\`{i}a-Bellido, D. G. Figueroa, Phys. Rev. Lett. {\bf 98}, 061302 (2007).
\bibitem{Dufaux_et_al}
J.-F. Dufaux, D. G. Figueroa, J. Garcia-Bellido, Phys. Rev. D {\bf 82}, 083518 (2010).
\bibitem{Abbott_et_al-1}
B. P. Abbott {\it et al.}, Astrophys. J. {\bf 909} 218 (2021). 
\bibitem{Abbott_et_al-2}
B. P. Abbott {\it et al.}, Astrophys. J. {\bf 923} 279 (2021).
\bibitem{Sakstein_et_al}
J. Sakstein and B. Jain, Phys. Rev. Lett. {\bf 119}, 251303 (2017).
\bibitem{Ezquiaga_et_al}
J. M. Ezquiaga and M. Zumalac\`{a}rregui, Phys. Rev. Lett. {\bf 119}, 251304 (2017).
\bibitem{Baker_et_al}
T. Baker, E. Bellini, P. G. Ferreira, M. Lagos, J. Noller, and I. Sawicki, 
Phys. Rev. Lett. {\bf 119}, 251301 (2017).
\bibitem{Pratten_et_al}
G. Pratten, P. Schmidt, and N. Williams, 
Phys. Rev. Lett. {\bf 129}, 081102 (2022). 
\bibitem{Corman_et_al}
M. Corman, A. Ghosh, C. Escamilla-Rivera, M. A. Hendry, S. Marsat, and N. Tamanini,
Phys. Rev. D {\bf 105}, 064061 (2022). 
\bibitem{Riess_et_al}
A. G. Riess et al.,
 Astron. J. {\bf 116} (1998), 1009.
\bibitem{Perlmutter_et_al}
S. Perlmutter et al., 
Astrophys. J. {\bf 517} (1999), 565.
\bibitem{Peebles}
P. J. E. Peebles and B. Ratra, Rev. Mod. Phys. {\bf 75}, 559 (2003).
\bibitem{Sahni}
V. Sahni, Class. Quantum Grav. {\bf 19} 3435 (2002).
\bibitem{brax}
P. Brax, Contemporary Physics {\bf 45}, 227 (2004).
\bibitem{Caldwell_et_al}
R. R. Caldwell, R. Dave and P. J. Steinhardt, Phys. Rev. Lett. {\bf 80} 1582 (1998) [astro-ph/9708069].
\bibitem{Zlatev_et_al}
I. Zlatev, L. M. Wang and P. J. Steinhardt, Phys. Rev. Lett. {\bf 82} 896 (1999) [astro-ph/9807002].

\bibitem{Armendariz-Picon_et_al-3}
C. Armendariz-Picon, T. Damour, and V. F. Mukhanov, 
Phys. Lett. B, {\bf 458}, 209 (1999).
\bibitem{Armendariz-Picon_et_al-1}
C. Armendariz-Picon, V. Mukhanov and P. J. Steinhardt, Phys. Rev. Lett. {\bf 85} 4438 (2000) [astro-ph/0004134].
\bibitem{Armendariz-Picon_et_al-2}
C. Armendariz-Picon, V. Mukhanov and P. J. Steinhardt, Phys. Rev. D {\bf 63} 103510 (2001) [astro-ph/0006373].
\bibitem{asm2}
N. Bose and A. S. Majumdar, Phys. Rev. D {\bf 79}, 103517 (2009).
\bibitem{asm3}
N. Bose and A. S. Majumdar, Phys. Rev. D {\bf 80}, 103508 (2009).
\bibitem{Bento_et_al}
M. C. Bento, O. Bertolami and A. A. Sen, Phys. Rev. D {\bf 66} 043507 (2002) [gr-qc/0202064].
\bibitem{Sen}
A. Sen, JHEP {\bf 07} (2002) 065 [hep-th/0203265].
\bibitem{Padmanabhan}
T. Padmanabhan, Phys. Rev. D {\bf 66} 021301 (2002) [hep-th/0204150].
\bibitem{Ali1}
Amna Ali, M. Sami, A.A. Sen, Phys.Rev.D {\bf 79}123501 (2009)

\bibitem{Dvali_et_al}
G. R. Dvali, G. Gabadadze and M. Porrati, Phys. Lett. B {\bf 485} 208 (2000) [hep-th/0005016].
\bibitem{asm1}
A. S. Majumdar, Phys. Rev D {\bf 64}, 083503 (2001).
\bibitem{Schwarz}
D. J. Schwarz, 
 18th IAP Colloquium on the Nature of Dark Energy: Observational and Theoretical Results on the Accelerating Universe, 
(2002) [astro-ph/0209584].
\bibitem{Rasanen}
S. R\"{a}s\"{a}nen, 
JCAP {\bf 02} (2004) 003 [astro-ph/0311257].
\bibitem{Wiltshire}
D.L. Wiltshire, 
6th International Heidelberg Conference on Dark Matter in Astro and Particle Physics, (2007), pp. 565-596, [arXiv:0712.3984].
\bibitem{Ali2}
Amna Ali, A. S. Majumdar,JCAP {\bf 01} 054 (2017).
\bibitem{Babichev_et_al-2}
E. Babichev, V. Dokuchaev, and Yu. Eroshenko, 
Phys. Rev. Lett. {\bf 93}, 021102 (2004). 
\bibitem{Babichev_et_al-1}
E.O. Babichev, V.I. Dokuchaev, and Y.N. Eroshenko,
J. Exp. Theor. Phys. {\bf 100}, 528-538 (2005).
\bibitem{Gao_et_al}
C. Gao, X. Chen, V. Faraoni, and Y. Shen,
Phys. Rev. D {\bf 78}, 024008 (2008).
\bibitem{Cheng-Yi}
Sun Cheng-Yi, 
 Commun. Theor. Phys. {\bf 52} 441 (2009).
\bibitem{Pepe_et_al}
C. Pepe, L. J. Pellizza, and G. E. Romero, 
Mon. Not. R. Astron. Soc. {\bf 420}, 3298-3302 (2012).  

\bibitem{Houghton_et_al}
L. Mersini-Houghton and A. Kelleher, 
Nuclear Physics B - Proceedings Supplements, Volume {\bf 194}, 272-277 (2009). 
\bibitem{Enander_et_al}
J. Enander and E. M\"{o}rtsell, 
Phys. Lett. B {\bf 683}, Issue 1, 7-10 (2010). 
\bibitem{Bondi}
H. Bondi, 
Mon. Not. R. Astron. Soc. {\bf 112}, 2 195-204 (1952). 
\bibitem{Petrich_et_al}
L. I. Petrich, S. L. Shapiro, R. F. Stark, and S. A. Tuekolsky,
Astrophys. J. {\bf 336} (1989).
\bibitem{string}
M. B. Green, J. Schwartz J and E. Witten,  {\it Superstring Theory} (Cambridge: Cambridge University Press, 1987).
\bibitem{string2}
M. Gasperini, F. Piazza and G. Veneziano, Phys. Rev. D {\bf 65} 023508 (2002). 
\bibitem{Piazza_et_al}
F. Piazza and S. Tsujikawa JCAP {\bf 07} (2004) 004.
\bibitem{Ohashi_et_al}
J. Ohashi and S. Tsujikawa, 
Phys. Rev. D {\bf 83}, 103522 (2011). 
\bibitem{Arkani-Hamed_et_al}
N. Arkani-Hamed, H. C. Cheng, M. A. Luty, and S. Mukohyama, 
JHEP {\bf 0405 } (2004) 074. 
\bibitem{Amendola_et_al}
L. Amendola and S. Tusjikawa, Dark energy - Theory and Observations, Cambridge University Press (2010). 
\bibitem{Erickson_et_al}
J. K. Erickson, R. R. Caldwell, P. J. Steinhardt, C. Armendariz-Picon, and V. Mukhanov, 
Phys. Rev. Lett. {\bf 88}, 12 (2002). 
\bibitem{catalog-1}
A. H. Nitz et al, 
Astrophys. J. {\bf 872} 195 (2019).
\bibitem{catalog-2}
B. P. Abbott et al, 
(LIGO Scientific Collaboration and Virgo Collaboration), 
Phy. Rev. X {\bf 9}, 031040 (2019).

\bibitem{Peters_et_al}
P. C. Peters and J. Mathews, Phys. Rev. {\bf 131} 435, (1963).  
\bibitem{M.Turner}
M. Turner, Astrophys. J. {\bf 216 }, 610-619, (1977).

\bibitem{arnab1}
A. Sarkar, K. R. Nayak, A. S. Majumdar, Phys, Rev D {\bf 100}, 103514 (2019).

\bibitem{Maggiore}
M. Maggiore, Gravitational waves: theory and experiments, Oxford University Press, New York (2008).
\end{thebibliography}
\end{document}